\documentclass[epsf]{elsart}

\usepackage{graphicx}
\usepackage{amssymb}

\journal{Physica D}

\begin{document}
\begin{frontmatter}

[Physica D {\bf 216}, 31 (2006)]

\title{Discrete Breathers in Two-Dimensional Anisotropic Nonlinear Schr\"odinger lattices}

\author[CONDMAT,ICMA,BIFI]{J. G\'omez-Garde\~nes}\ead{gardenes@unizar.es},
\author[CONDMAT,ICMA,BIFI]{L.M. Flor\'{\i}a} and
\author[LANL]{A.R. Bishop}

\address[CONDMAT]{Dpt. de F\'{\i}sica de la Materia Condensada,
Universidad de Zaragoza, 50009 Zaragoza, Spain.}
\address[ICMA]{Dept. de Teor\'{\i}a y Simulaci\'on de Sistemas Complejos,
Instituto de Ciencia de Materiales de Arag\'on (ICMA), C.S.I.C.-Universidad
de Zaragoza, 50009 Zaragoza, Spain.}
\address[BIFI]{Instituto de Biocomputaci\'on y F\'{\i}sica de Sistemas
Complejos (BIFI), Universidad de Zaragoza, Zaragoza 50009, Spain}
\address[LANL]{Theoretical Division and Center for Nonlinear
Studies, Los Alamos National Laboratory, Los Alamos, New Mexico~87545}

\begin{abstract}

We study the structure and stability of discrete
breathers (both pinned and mobile) in two-dimensional nonlinear 
anisotropic Schr\"odinger lattices. Starting from a 
set of identical one-dimensional systems we develop
the continuation of the localized pulses from the weakly coupled
regime (strongly anisotropic) to the homogeneous one (isotropic). 
Mobile discrete breathers are seen to be a superposition of a localized
mobile core and an extended background of two-dimensional
nonlinear plane waves. This structure is in agreement with 
previous results on one-dimensional breather mobility. The study
of the stability of both pinned and mobile solutions is performed
using standard Floquet analysis. Regimes of quasi-collapse 
are found for both types of solutions, while
another kind of instability (responsible for the discrete
breather fission) is found for mobile solutions. The development of
such instabilities is studied examining typical trajectories on the 
unstable nonlinear manifold. 
\end{abstract}

\end{frontmatter}

\section{Introduction}
\label{sec:Introduction}

We present a numerical study of exact breather dynamics in
two-dimensional nonlinear Schr\"odinger lattices. 
This issue fits well with the enduring scientific 
interests of Serge Aubry in the many faceted subject 
of localization and transport in nonlinear macroscopic discrete
systems, where Serge's outstanding contributions are widely
recognized. Many ideas and lines of study followed in the
investigation reported here have found a source of inspiration in
early works of Aubry and collaborators, among which the doctoral works
of Th. Cretegny and J.L. Mar\'{\i}n deserve special mention.

The existence and properties of localized solutions in extended
discrete systems have atracted interest in a broad range of
physical fields \cite{Scott}. Discrete breathers,
sometimes referred to as intrinsic localized modes, are
spatially localized and time periodic solutions. 
These solutions arise in the context of nonlinear discrete systems and
are of fundamental interest for varied physical applications such as pulse
propagation in nonlinear optics, energy storing and transport in
biomolecules, plasma physics, etc... The existence of discrete breathers in these systems has been 
proved rigorously \cite{Mackay-Aubry} for a number of equations 
with physical relevance and, contrary to  continuous
nonlinear equations, their existence can be regarded
as a generic feature of these systems.
One of the most important class of equations are the so-called 
{\em discrete nonlinear Schr\"odinger lattices} \cite{Escorial,Kevrekidis-IJMPB01}. 
The existence of discrete breathers has been proven for a wide range of
systems belonging to this class of nonlinear difference-differential
equations. In particular, the most important example of wide aplicability is
the standard nonlinear Schr\"odinger equation. For instance, 
this equation was employed in \cite{Aceves1,Aceves2} for describing 
the propagation of localized beams in an array of nonlinear (Kerr
type) waveguides, having experimental validation subsequently 
reported in \cite{Eisemberg,Morandotti}.   

The study of two-dimensional nonlinear Schr\"odinger lattices has
atracted much atention \cite{Kev-Ras-PRE00-1,Kev-Ras-PRE00-2} in recent 
years due to the new phenomena emerging when the dimensionality of 
the lattice is increased. Some examples of these new features are 
the existence of vortex-breathers \cite{Mal-Kev-PRE01} which supports
energy flux, the appearance of an energy threshold for the creation 
of discrete breathers
\cite{Flach,Weinstein,Kastner1,Kastner2,Kalosakas} 
and the ubiquity of an instability (the quasi-collapse) of some 
discrete breather solutions leading to a highly localized pulson
state \cite{Mezent-Mush-JETP94,Laedke-PRL94,Laedke-JETP95,
Rasmussen_Rygdal,Christ-Gaid-PRB96,Christiansen-PScr}. 
These theoretical efforts have their counterpart in recent advances in
the field of nonlinear optics. The studies of two-dimensional arrays
of coupled nonlinear waveguides allow the experimental observation of those
effects studied theoretically. Specially relevant is the recent experimental
breakthrough (theoretically designed in \cite{Efremidis}) 
by Fleisher {\em et al} \cite{Fleisher1,Fleisher2},
where a two-dimensional array of nonlinear waveguides is induced in a
photosensitive material. This technique provides a clear experimental
verification of the two-dimensional discrete breather existence in this
system. In particular, besides the observation of standard discrete breathers, 
these works reported the first observations of staggered discrete
breathers.

Our study here will focus on the computation of numerically exact dicrete
breathers in two-dimensional anisotropic nonlinear Schr\"odinger
lattices, {\em i.e.} where the couplings in the two spatial directions
are different.  The use of shooting methods allow us to find these
solutions and analyze their structural and stability properties. Both
pinned and mobile discrete breathers are studied. In the latter case we
will study only the ones whose motion is along one axis of the
lattice. The analysis of the numerically exact solutions shed light
on some features of the properties and stability of localized solutions 
reported in previous works. 

The plan of the paper is as follows: In section
\ref{sec:Numerical} we provide the technical background and details
needed for self-contained purposes. First we summarize in 
\ref{subsec:1D} the main conclusions on the
dynamics of 1D Schr\"odinger discrete breathers reported in
\cite{JGG1,JGG2}. A detailed account of the numerical methods
(SVD-regularized Newton continuation of operator fixed points)
that we have used can be found in that reference. Also in
\ref{subsec:1D}, we discuss briefly the most relevant formal
differences with respect to alternative approaches to
(one-dimensional) exact mobility of discrete breathers, {\em e.g.} those
in refs \cite{Musslimani} and \cite{others,others2}. In subsection
\ref{sec:Numerics} we introduce the two-dimensional anisotropic
Salerno lattice and provide explanations on the implementation of
the numerical procedures used to study the dynamics of 2D discrete
breathers.

The analysis of the results of our numerics on pinned discrete
breathers for anisotropic nonlinear Schr\"odinger lattices is reported in
section \ref{sec:pinned}. We present the numerical computations of
the fixed point norm, as a function of three parameters: breather frequency, 
transversal coupling, and nonlinearity (see below). They show, as anticipated,
the so-called {\em quasi-collapse} transition, associated with the
(well-known) existence of thresholds for the
breather norm in two-dimensional lattices. We present numerically
computed sectors of the bifurcation surface. 
We conclude this section with a brief look at the nonlinear
dynamics on the unstable manifold, whose typical trajectories have
been called {\em pulson} states. Early numerical work on the 2D
quasi-collapse phenomena in isotropic lattices was reported in
\cite{Christ-Gaid-PRB96,Christiansen-PScr} and
\cite{Laedke-JETP95}. A two-year-old account of
the "state of knowledge" on 2D Schr\"odinger lattices can be found
in Section six of \cite{Escorial}.

In section \ref{secc:mobile} we show results on a type of mobile
breather, namely those moving along the direction of
stronger lattice coupling constant. The structure of each of
these mobile exact discrete breathers is that of a localized
moving {\em core} superimpossed on a specific extended state of 
resonant small amplitude radiation, the
{\em background}. We present here the results of an extensive
Floquet stability analysis of this type of solutions in two
sectors of the three-dimensional parameter space, 
which clearly show the existence of two different
transitions. The tangent space eigenvectors associated to each of
the transitions are presented, and the relation of the unstable manifold
trajectories to pulson states is analyzed afterwards.

We conclude with section \ref{secc:conclusions}, where we
briefly review the results obtained and illustrate their possible
implications for mobility of 2D discrete breathers.

\section{Numerical continuation and preliminary results}
\label{sec:Numerical}

The use of numerical tools for the continuation of discrete breather
solutions has been widely employed since their existence proof was
reported (see {\em e.g} \cite{Marin-Aubry,Tesis-Marin}). 
In particular, the design of numerical techniques for finding exact mobile breathers
based on those employed for the pinned ones has been explored in the recent
years
\cite{JGG1,JGG2,Aubry-Creteg-PhysD98,Flach-Kladko-PhysD99,Sanchez-Brey}
and paves the way to resolving the still open question
about discrete breather mobility. Here, after reviewing the
most important features of 1D mobile breathers in nonlinear
Schr\"odinger lattices, we briefly explain how the continuation method 
is implemented in our system. 

\subsection{Mobility of one-dimensional nonlinear Schr\"odinger discrete breathers}
\label{subsec:1D}

As was introduced in section \ref{sec:Introduction}, exact
mobility of dicrete breathers in 1D Schr\"odinger
nonlinear lattices were numerically studied by the authors in 
previous works \cite{JGG1,JGG2}. In particular, a numerical 
continuation from the integrable {\em Ablowitz-Ladik lattice}, 
A-L, (where exact mobile breathers can be calculated analytically 
\cite{AL}), 
\begin{equation}
{\mbox i} \dot{\Phi}_n= -(\Phi_{n+1} + \Phi_{n-1})\left[ 1 + \frac{\gamma}{2}
|\Phi_n|^2 \right]\;,
\label{eq:AL}
\end{equation}
to the standard {\em discrete nonlinear Schr\"odinger equation}, DNLS,
\begin{equation}
{\mbox i} \dot{\Phi}_n= -(\Phi_{n+1} + \Phi_{n-1})
- \gamma |\Phi_n|^2 \Phi_n\;,
\label{eq:DNLS}
\end{equation}
was performed within the so-called {\em Salerno model} \cite{Salerno},
\begin{equation}
{\mbox i} \dot{\Phi}_n= -(\Phi_{n+1} + \Phi_{n-1})\left[ 1 + \mu
|\Phi_n|^2 \right] - 2 \nu \Phi_n |\Phi_n|^2\;,
\label{eq:Salerno}
\end{equation}
where $\gamma$, $\mu$ and $\nu$ are the parameters accounting for the 
strength of the nonlinear terms. The above equation includes the former 
two relevant equations (\ref{eq:AL}) and (\ref{eq:DNLS}), for 
($\mu=\gamma/2$, $\nu=0$) and ($\mu=0$, $\nu=\gamma/2$) respectively, 
providing the desired interpolation needed to develop the continuation 
scheme. The integrability of equation (\ref{eq:AL}) provides for $\gamma>0$ a 
two-parametric family of discrete breathers
\begin{equation}
\Phi_{n}(t)=\sqrt{\frac{2}{\gamma}}\;{\mbox{sinh}}\beta\;{\mbox{sech}}[\beta(n-x(t))]\;{\mbox{exp}}[{\mbox{i}}(\alpha(n-x(t))-\Omega(t))]\;,
\end{equation}
where the two parameters $\alpha\in[-\pi:\pi]$ and $\beta>0$ describe the velocity and internal frequency of the solution
\begin{eqnarray}
v_b&=&\dot{x}=2\;{\mbox{sinh}}\beta\;{\mbox{sin}}\alpha/\beta\;,
\\
\omega_b&=&\dot{\Omega}=2\;{\mbox{cosh}}\beta\;{\mbox{cos}}\alpha+\alpha v_b\;.
\end{eqnarray}
The Salerno lattice (\ref{eq:Salerno}), possess two dynamical invariants,
namely the hamiltonian,
\begin{eqnarray}
{\mathcal{H}}= &-&\sum_{n}(\Phi_{n}\overline{\Phi}_{n+1} +
\overline{\Phi}_{n}\Phi_{n+1}) -2\frac{\nu}{\mu}\sum_{n}
|\Phi_n|^2 \nonumber
\\
&+&2\frac{\nu}{\mu^2}\sum_{n}\ln(1+\mu|\Phi_n|^2)\;,
\label{eq:Salerno_Ham}
\end{eqnarray}
where $\overline{\Phi}_{n}$ denotes the complex conjugate of
${\Phi}_{n}$, and the norm
\begin{equation}
{\mathcal{N}} = \frac{1}{\mu}\sum_{n}\ln(1+\mu|\Phi_n|^2)\;.
\label{eq:Salerno_Norm}
\end{equation}

In order to find exact mobile discrete breathers in the Salerno
model we define a ($p,q$)-resonant solution $\Phi_{n}(t)$ referred
to some time scale $\tau$ such that
\begin{equation}
\Phi_{n}(t_{0})=\Phi_{n+p}(t_{0}+q\tau)\;.
\label{eq:p-q_resonant}
\end{equation}
Within the above definition, a mobile breather that translates
$p$ sites after $q$ periods of the internal oscillation will
satisfy equation (\ref{eq:p-q_resonant}) when $\tau=T_{b}$. In
this sense, the continuation focuses on the families of
($p,q$)-resonant discrete breathers, that is breather solutions
with the two characteristic time scales (corresponding to the
breather velocity, $v_{b}$, and the internal frequency,
$\omega_{b}$) being commensurate. In all the computations we have 
used finite lattices with periodic boundary conditions (PBC) so 
that $\Phi_{N+1}=\Phi_{1}$ and $\Phi_{0}=\Phi_{N}$ (with $N$ being the 
lattice size).

In the previous works \cite{JGG1,JGG2} the authors start from 
those A-L solutions which are ($p,q$)-resonant, 
we discretize (fine grid) the path $\mu+\nu=1$ (with $\mu$ and $\nu$ 
being positive) along the Salerno model, and go through 
it computing the corresponding ($p,q$)-resonant solutions 
for the pairs $(\mu,\nu)$. We can choose the path without loss 
of generality because of the scaling property of equation 
(\ref{eq:Salerno}). Then, each solution is numerically computed as a 
fixed point of the map
\begin{equation}
M = L^{p} {T}_{(\omega_{b},\nu)}^{q}\;,
\end{equation}
where ${L}$ is the lattice translation operator
${L}(\{\Phi_n(t_{0})\})=\{\Phi_{n+1}(t_{0})\}$, and
${T}_{(\omega_{b},\nu)}$ is the $T_b$-evolution map ($T_b = 2\pi
/\omega_b$) following the dynamics dictated by equation
(\ref{eq:Salerno}) for the corresponding value of $\nu$
($\mu=1-\nu$); {\em i.e.},
${T}_{(\omega_{b},\nu)}[\{\Phi_n(t_{0})\}]=\{\Phi_{n}(t_{0}+T_b)\}$.

The continuation was then performed for a fine grid of frequencies
belonging to the family of ($p=1,q=1$)-resonant discrete
breathers. The most important conclusion about discrete breathers
in these nonlinear Schr\"odinger lattices is that mobility in the
non-integrable regime ($\nu\neq1$) demands the presence of an
extended {\em background} to which the fixed point solution 
is spatially asymptotic ($n \rightarrow \infty$),
{\em i.e.} the solution is exactly written as
\begin{equation}
\Phi_{n}(t)=\Phi_{n}^{{\mbox {core}}}(t)+\Phi_{n}^{{\mbox {bckg}}}(t)\;.
\end{equation}
This expression defines the purely localized component
$\Phi_{n}^{{\mbox {core}}}(t)$ of the solution.
The background is a finite linear combination of nonlinear
plane waves, $\Phi_{n}^{{\mbox {bckg}}}(t)=\sum_{j=1}^{s}
A_{j}{\mbox exp}[{\mbox i}(k_{j}n-\omega(k_{j},A_{j}) t)]$. These plane 
waves are exact solutions of the Salerno model (\ref{eq:Salerno}) being the 
nonlinear dispersion relation
$\omega(k_{j},A_{j}) = -2 [1 + \mu |A_{j}|^{2}]\cos k_{j} - 2\nu |A_{j}|^{2}$.
The results concerning the characterization of the
background are discussed in detail in \cite{JGG1,JGG2}, here we briefly 
summarize the most important features:
\begin{itemize}
\item[{\it (i)}] The set of ``$s$'' plane waves which take part in the 
background of a $(p,q)-$resonant discrete breather with internal frequency 
$\omega_b$ is derived by the simple selection rule for the wave-numbers $k_{j}$
\begin{equation}
\frac{\omega(k_{j},A_{j})}{\omega_b} = \frac{1}{q}\left(\frac{p}{2\pi} k_{j} -m
\right)\;,
\label{resonant}
\end{equation}
{\em i.e.} only the plane waves which are ($p,q$)-resonant with
the internal period of the breather can be components of
$\{\Phi_{n}^{{\mbox{bckg}}}(t)\}$. The number of solutions of 
(\ref{resonant}) fixes ``$s$''.

\item[{\it (ii)}] The amplitudes $\{ A_{j}\}$ of the nonlinear
plane waves differ by orders of magnitude.

\item[{\it (iii)}] There exist a strong positive correlation
between the amplitude of the background and the strength of the
Peierls-Nabarro barrier arising from the periodic lattice. 
This correlation is particularly clear when
symmetry breaking transitions occur for the also studied case of 
$\nu<0$ and $\mu>0$ (see reference \cite{JGG2}),
and reflects the link between non-integrability and the existence
of the background dressing of the mobile core.

\item[{\it (iv)}] Finally, the interpretation of the correlation
described in {\it (iii)} is reinforced from a study of the energy
evolution of the mobile core: {\em There is an energy balance brought by
the background when the core moves along the lattice}. In particular,
it can be observed how the core energy oscillates periodically so that
it takes the maximum energy value when the core visits the inter-site
configuration. This extra energy periodically obtained by the core is
provided by the background, with the energy maximum clearly related to the
background amplitude.
\end{itemize}

Before concluding this subsection, it is worth commenting on some
of the differences between the Newton continuation
of fixed points that we use in this paper, and other important 
recent approaches to breather numerics. 
The work by Ablowitz {\em et al} \cite{Musslimani} uses discrete Fourier
analysis to obtain a nonlinear nonlocal integral equation, from
where the " ... soliton is thus viewed as a fixed point of a
nonlinear functional" (sic) in the Fourier transformed space of
functions. Following these authors, their results seem to differ
from those of early pioneering work \cite{Feddersen} (nowadays
textbook material \cite{Scott}) "in which a {\em continuous}
travelling solitary waves were reported using Fourier series
expansions with finite period $L$ while assuming convergence as
$L\rightarrow \infty$" (sic). Ablowitz {\em et al} term {\em
continuous} a solution that can be defined off the lattice
points, which they see as "necessary when discussing travelling
waves in lattices" (sic), and disagree with some conclusions
reported in the earlier works.

The ("orthodoxy matters") discussion above helps us to clarify how
differently our numerical approaches "sees" the discrete
Schr\"odinger breather problem: The very concept of a variable
defined off the lattice points is intrinsically alien to our
discrete approach, which neither needs of it nor excludes its
eventual consideration. In contrast to those views (but not at all
in logical opposition), we consistently view the thermodynamical
limit ($N\rightarrow \infty$) in lattice space, much in the sense 
used {\em e.g.} by Serge Aubry in his celebrated
work on the Frenkel-Kontorova ground state problem \cite{Aubry}:
The infinite size limit is built up from a subsequence of PBC 
(finite) lattices for which the limit is well defined. This will
make the Fourier-transformed $k$-space continuum. 

Closer to our approach in some respects, though technically
different in many others, is the formal approach purposed recently
by James and collaborators \cite{others,others2}. This
approach, which uses recent central manifold theorems, was brought
to our attention very recently and
unfortunately, we have to defer to a future publication a 
comparison of our 1D computations with theirs, this 
is beyond the scope of this paper, namely,
the anisotropic 2D Sch\"rodinger lattices, which we introduce in
the next subsection.

\subsection{Two-dimensional anisotropic lattices.}
\label{sec:Numerics}

Motivated by the method described above, we can extend the
continuation scheme for calculating exact discrete breathers in
higher dimensional systems. In particular we focus on the
two-dimensional nonlinear Schr\"odinger lattice
\begin{eqnarray}
{\mbox i}\dot{\Phi}_{nm}=
&-&C_{1}(\Phi_{n+1,m}+\Phi_{n-1,m})(1+\mu|\Phi_{n,m}|^2)
\nonumber
\\
&-&C_{2}(\Phi_{n,m+1}+\Phi_{n,m-1})(1+\mu|\Phi_{n,m}|^2)
\nonumber
\\
&-&2\nu\Phi_{n,m}|\Phi_{n,m}|^2\;.
\label{eq:2D-DNLS}
\end{eqnarray}

This lattice can be viewed as the  {\em two-dimensional Salerno
model}. The two coupling parameters $C_{1}$ and $C_{2}$ provide a
technical advantage for numerics (see below), but they are also
introduced for theoretical and experimental interest. The
possibility of controlling the ratio between the two linear couplings
of the two transversal directions has been studied in various works
as a way of analysing how the intrinsic 2D phenomena 
(such as the quasi-collapse) emerge. In fact, for $C_{1}<<C_{2}$, $\mu=0$
and $\nu=\gamma/2$  
equation (\ref{eq:2D-DNLS}) describes a set of weakly coupled
nonlinear wavequide arrays and can be considered as a case 
of ``{\em intermediate dimensionality}''. This extreme has been 
studied experimentally in \cite{Cheskis} and using perturbative
methods in \cite{Lederer}. On the other hand, this equation
incorporates, as two particular limits, the physically relevant 
standard two-dimensional DNLS equation ($\mu=0$, $\nu=\gamma/2$) and the
two-dimensional A-L lattice ($\mu=\gamma/2$, $\nu=0$). 
The continuation between these two limits provides a useful tool 
for studying the interplay between the on-site and inter-site 
nonlinearity in the 2D anisotropic lattice. Moreover, the anisotropy 
(or freedom in the values of the coupling parameters $C_1$ and $C_2$) 
allows to include an integrable model among the members of the family 
of nonlinear lattices described by (\ref{eq:2D-DNLS}). That is, for $\nu=0$, 
$C_{i}=0$ and $C_{j}\neq0$ one obtains a set of integrable A-L chains.
In this way, every 2D model included in (\ref{eq:2D-DNLS}) is connected 
with an integrable model where analytic discrete breathers are available.
In what follows we set $\gamma=2$ and $\nu=1-\mu$ with $\nu>0$ and $\mu>0$.

The dynamics (\ref{eq:2D-DNLS}) can be derived from the Salerno 
Poisson structure 
\begin{equation}
\left\{{\mathcal A}, {\mathcal B}\right\}=\sum_{n,m}\left(\frac{\partial
{\mathcal A}}{\partial \Phi_{n,m}}\frac{\partial
{\mathcal B}}{\partial \overline{\Phi}_{n,m}}-\frac{\partial
{\mathcal A}}{\partial \overline{\Phi}_{n,m}}\frac{\partial
{\mathcal B}}{\partial \Phi_{n,m}}\right)\cdot\left(1+\mu|\Phi_{n,m}|^2\right)\;,
\label{eq:2DPoisson}
\end{equation}
with the hamiltonian
\begin{eqnarray}
{\mathcal H}=
&-&C_{1}\sum_{n,m}\left(\Phi_{n,m}\overline{\Phi}_{n+1,m}+\Phi_{n+1,m}\overline{\Phi}_{n,m}\right)
\nonumber
\\
&-&C_{2}\sum_{n,m}\left(\Phi_{n,m}\overline{\Phi}_{n,m+1}+\Phi_{n,m+1}\overline{\Phi}_{n,m}\right)
\nonumber
\\
&-&\frac{2\nu}{\mu}\sum_{n,m}|\Phi_{n,m}|^2
+\frac{2\nu}{\mu^2}\sum_{n,m}{\mbox{ln}}\left(1+\mu|\Phi_{n,m}|^2\right)\;.
\label{eq:2DHamiltonian}
\end{eqnarray}
As in the 1D Salerno model there is also a second conserved
quantity, the norm, due to the phase invariance of the equations of motion 
(\ref{eq:2D-DNLS})
\begin{equation}
{\mathcal N}=\frac{1}{\mu}\sum_{n,m}{\mbox{ln}}\left(1+\mu|\Phi_{n,m}|^2\right)\;.
\label{eq:2DNorm}
\end{equation}

In the same manner as in the 1D case we will focus on a special set
of 2D discrete breathers. For this, we have to generalize the 
definition (\ref{eq:p-q_resonant}) of a resonant solution in 
the two-dimensional case. In this
context discrete breathers solutions are characterized by three
time scales. Namely, one associated with the internal
oscillation $\omega_{b}$ and the other two derived from the
translation of the localization center, {\em i.e.} its velocity
$\vec{v}_{b}=(v_{x},v_{y})$. The subset of $3$-tuples
($\omega_{b}$,$\vec{v}_{b}$) that fulfil the
($p_{x},p_{y},q$)-resonance condition
\begin{eqnarray}
v_{x}\frac{2\pi}{\omega_{b}}&=&\frac{p_{x}}{q}
\\
v_{y}\frac{2\pi}{\omega_{b}}&=&\frac{p_{y}}{q}\;,
\label{eq:p-qRES}
\end{eqnarray}
(where $p_{x}$, $p_{y}$ and $q$ are integers) denote the breather
solution that can be obtained with our continuation method. These
solutions are those that after $q$ periods of the internal
frequency, $\hat\Phi(t_{0}+qT_{b})$, translates $p_{x}$ and $p_{y}$
lattice sites in the $x$ and $y$ direction of the square lattice,
respectively, {\em i.e.}
\begin{equation}
\hat\Phi_{n,m}(t_{0})=\hat\Phi_{n+p_{x}, m+p_{y}}(t_0 +qT_{b})\;,
\end{equation}
where again PBC are applied $\Phi_{N_{x}+1,m}=\Phi_{1,m}$, 
$\Phi_{0,m}=\Phi_{N_{x},m}$, $\Phi_{n,N_{y}+1}=\Phi_{n,1}$ 
and $\Phi_{n,0}=\Phi_{n,N_{y}}$ (with $N_{x}$ and $N_{y}$ 
being the lattice size in the $x$ and $y$ direction respectively).
Consequently, a ($p_{x},p_{y},q$)-resonant state
will be a solution of the following set of equations
\begin{eqnarray}
F_{(p_{x},p_{y},q,\omega_{b},\nu, C_{1}, C_{2})}[\{\hat{\Phi}_{n,m}(t_{0})\}]
&=&{L}_{y}^{p_{y}}{L}_{x}^{p_{x}}
{T}_{(\omega_{b}, \nu, C_{1}, C_{2})}^q[\{\hat{\Phi}_{n,m}(t_{0})\}]=
\nonumber
\\
&=&\{\hat{\Phi}_{n,m}(t)\}\;,
\label{eq:2D-Map}
\end{eqnarray}
where the operators ${L}_{i}$ are the lattice translation in
the $i$-direction, 
\begin{eqnarray}
L_{x}[\{\Phi_{n,m}(t_{0})\}]&=&\{\Phi_{n+1,m}(t_{0})\}\;,
\\
L_{y}[\{\Phi_{n,m}(t_{0})\}]&=&\{\Phi_{n,m+1}(t_{0})\}\;.
\end{eqnarray}
Besides, ${T}_{(\omega_{b},\nu,C_{1},C_{2})}$ is the time evolution 
operator given by equation (\ref{eq:2D-DNLS}) over one period 
$T_{b}=2\pi/\omega_{b}$,
\begin{equation}
{T}_{(\omega_{b},\nu,C_{1},C_{2})}[\{\Phi_{n,m}(t_{0})\}]=
\{\Phi_{n,m}(t_{0}+T_{b})\}\;.
\end{equation}

\begin{figure}[!tbh]
\begin{tabular}{cc}
\centerline{
\resizebox{14.cm}{!}{%
\includegraphics[angle=-0]{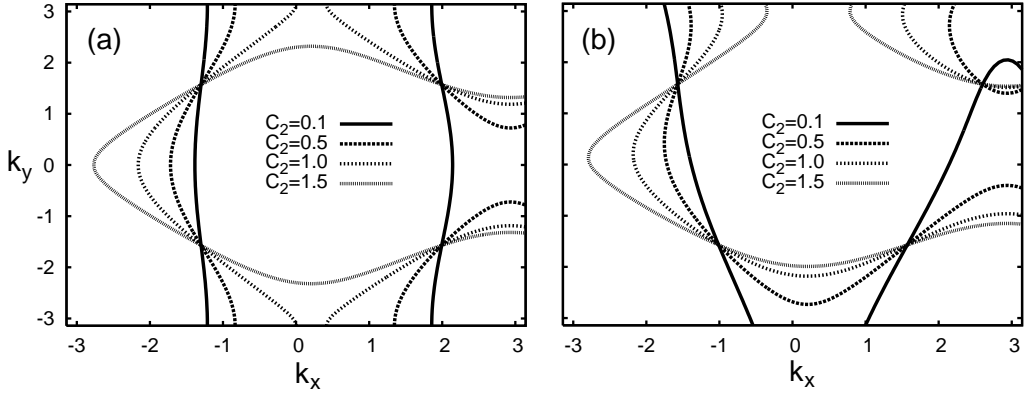}
}
}
\end{tabular}
\caption{Wave numbers, $\vec{k}=(k_{x},k_{y})$, of the
$(1,0,1)$ {\bf (a)} and $(1,1,1)$
{\bf (b)} resonant plane waves for $m=0$ (see
equation (\ref{eq:2D-SR})). Different values of $C_{2}$, while $C_{1}$ is
fixed ($C_{1}=1$), are shown. The reference time scale for the
resonance is set to $\tau=2.4315$ ($\omega=2.584$).}
\label{fig:resonances}
\end{figure}

In order to illustrate the 2D time scales resonance we
consider the plane wave solutions of equation (\ref{eq:2D-DNLS}):
$\Phi_{n,m}(t)=A\exp[{\mbox i}(k_{x}n+k_{y}m-\omega t)]$.
These solutions possess the following nonlinear dispersion relation
\begin{equation}
\omega(\vec{k},A)=2(C_{1}\cos k_{x}+C_{2}\cos k_{y})(1+\mu A^2)-2\nu A^2\;.
\label{eq:2D-DR}
\end{equation}
Hence, we can obtain the subset of plane waves which are
($p_{x},p_{y},q$)-resonant with some time scale $\tau$ ({\em
i.e.} after a time $q\tau$ they have translated $p_{x}$ and
$p_{y}$ sites in the $x$ and $y$ direction, respectively). Each
member of these subsets will be labelled by the pair
$\vec{k}=$($k_{x}$, $k_{y}$) and from the condition
(\ref{eq:2D-Map}) it follows that the corresponding set of values
of $\vec{k}$ for each family will satisfy the relation
\begin{equation}
\omega(\vec{k},A)=\frac{1}{q\tau}\left(\vec{p}\cdot\vec{k}-\frac{m}{2\pi}\right)\;,
\label{eq:2D-SR}
\end{equation}
where $m$ is an integer and $\vec{p}=$($p_{x}$, $p_{y}$). In figure
\ref{fig:resonances} the corresponding values of
$\vec{k}$ are represented for two resonances of the type
$(p_{x}=1,p_{y}=0,q=1)$ and $(p_{x}=1,p_{y}=1,q=1)$.

The method used for solving equation (\ref{eq:2D-Map}) 
for each resonant $3$-tuple ($\omega_{b}$,$\vec{v}_{b}$) 
is the same as in the 1D case, which is extensively described in \cite{JGG2}. 
The implicit function theorem assures that a fixed point
solution of a map (\ref{eq:2D-Map}) given by $\vec{\xi}=$($p_{x}$,
$p_{y}$, $q$, $\omega_{b}$, $\nu$, $C_{1}$, $C_{2}$) can be
obtained provided that {\it(i)} the Jacobian of the operator
$F_{\vec{\xi}}[\{\Phi_{n,m}(t_{0})\}]-{I}$ is invertible and {\it(ii)}
we know a fixed point of a map corresponding to an infinitesimally
close set of parameters, $\vec{\xi}-\delta\vec{\xi}=$ ($p_{x}$,
$p_{y}$, $q$, $\omega_{b}-\delta\omega_{b}$, $\nu-\delta\nu$,
$C_{1}-\delta C_{1}$, $C_{2}-\delta C_{2}$). The first demand can
be satisfied using a {\em singular value decomposition} (SVD)
\cite{Strang,Aubry-Creteg-PRB97,Tesis-Cretegny} of the Jacobian 
in order to obtain the pseudo-inverse operator. On the other hand, 
when the second condition is fulfilled convergence 
of the Newton-Raphson iterative scheme is guaranteed. For this, 
we start with a sufficiently good trial solution, 
$\{{\Phi}^{0}_{n,m}(t_{0})\}$ and solve the equation
\begin{equation}
\{\delta\Phi^{0}_{n,m}(t_{0})\}=
-DF_{\vec{\xi}}[\{\Phi^{0}_{n,m}(t_{0})\}]^{-1}\cdot F_{\vec{\xi}}[\{\Phi^{0}_{n,m}(t_{0})\}]\;,
\label{eq:Iteration}
\end{equation}
in order to obtain $\{\Phi^{1}_{n,m}(t_{0})\}=\{\Phi^{0}_{n,m}(t_{0})\}+\{
\delta\Phi^{0}_{n,m}(t_{0})\}$. We iterate this calculations to the
desired convergence, and then the solution
$\{\hat{\Phi}_{n,m}(t_{0})\}$, is obtained. In our numerics this is
the case when
\begin{equation}
F_{\vec{\xi}}[\{\Phi^{i}_{n,m}(t_{0})\}]<N\cdot10^{-16}\;,
\end{equation}
(where $N$ is the total number of sites in the square lattice) is
fulfilled. Once the solution is found we use it as the following trial solution,
$\{{\Phi}^{0}_{n,m}(t_{0})\}$, for solving the map (\ref{eq:2D-Map})
corresponding to the next set of parameters 
$\vec\xi^{\;'}=\vec\xi+\delta\vec{\xi}$.

As an additional benefit, this method provides the linear stability of the computed
solution. For this, we only have to compute the eigenvalues of the
Jacobian of the operator around the solution
$DF_{\vec\xi}[\{\hat{\Phi}_{n,m}(t_{0})\}]$. In fact, this Jacobian is the
extended Floquet matrix of the solution. Then, a solution is 
linearly stable if all the eigenvalues of the corresponding 
Floquet matrix are inside the unit circle. Moreover, the 
symplectic character of the dynamics (\ref{eq:2D-DNLS}) 
implies that all the eigenvalues appear in quadruplets 
($\lambda$, $\overline{\lambda}$, $1/\lambda$, $1/\overline{\lambda}$) 
and thus for a linearly stable solutions all the eigenvalues 
of its extended Floquet matrix lie on the unit circle.

There are two possible paths for developing the continuation
method depending on the choice of the starting point of the 
continuation. One possibility is to start from the full
anti-continuum limit, $C_{1}=C_{2}=0$, where a pinned breather
solution of frequency $\omega_{b}$ is written as
\begin{equation}
\hat{\Phi}_{n,m}(t)=\delta_{n,n_{0}}\delta_{m,m_{0}}
\sqrt{\frac{\omega_{b}}{2\nu}}\exp({\mbox i}\omega_{b}t)\;.
\label{eq:Sol_AC}
\end{equation}
Starting from the above solution, we can perform the continuation
increasing the parameters $C_{1}$ and $C_{2}$ as usual, and so
obtain the whole family of ($p_{x}=0,p_{y}=0,q=1$)-resonant
discrete breathers. An alternative path starts from the
one-dimensional limit, $C_{2}=0$. The choice of this second limit (which implies taking 
as the very initial trial solution of the continuation 
the whole set of 1D solutions described in section \ref{subsec:1D}) 
is justified when seeking mobile solutions. As stated above, 
this limit offers the possibility of studying strongly anisotropic 
lattices as a controlled interpolating situation between one and two
dimensions. On the other hand, employing this strategy we can only obtain
those solutions which are ($p_{x}=p,p_{y}=0,q$)-resonant,
{\em i.e.} the two-dimensional continuation of those
one-dimensional ($p=p_{x},q$)-resonant discrete breathers. Hence,
the solution from which we start is
\begin{equation}
\hat{\Phi}_{n,m}(t)=\delta_{m,m_{0}}\hat{\Phi}^{1D}_{n}(t)\;,
\label{eq:Sol_AC}
\end{equation}
where $\hat{\Phi}^{1D}_{n}(t)$ is the corresponding
($p=p_{x},q$)-resonant one-dimensional solution.

In what follows we will employ both continuation paths when we study
the case of pinned breathers (section \ref{sec:pinned}), and we will
show that the results obtained are the same when approaching the
same limit (the standard two-dimensional DNLS).

\section{Two-dimensional pinned discrete breathers}
\label{sec:pinned}

As we have discussed, we can choose two different starting points 
for the continuation of $(0,0,1)$-resonant fixed points (pinned 
breathers) of equation (\ref{eq:2D-Map}): {\it(i)} the full 
anti-continuum (AC) limit ($C_1=C_2=0$), or {\it (ii)} the 
(one-dimensional, 1D) limit of uncoupled chains ($C_1\neq 0, C_2=0$), 
where they can be obtained from continuation along increasing 
values of the parameter $\nu$ from the 1D A-L lattice
(\ref{eq:AL}). As a test for our codes, we have 
checked that both paths arrive to the same solution. 
In fact, unique continuations can proceed along any path 
on the plane of parameters ($C_2,\nu$) that we have explored.

Early works \cite{Mezent-Mush-JETP94,Laedke-PRL94,Laedke-JETP95} 
on the isotropic two-dimensional standard 
DNLS equation analyzed the so-called {\em quasi-collapse} instability 
of pinned discrete breathers, {\em i.e.} the condensation of all 
the energy into a few modes in discrete nonlinear systems, which 
corresponds to the onset of a singularity (wave collapse) 
\cite{Rasmussen_Rygdal} in multidimensional continuum models. 
Subsequent numerical works \cite{Christ-Gaid-PRB96} extended 
these studies to the isotropic 2D Salerno lattice and addressed 
the question of how the instability is affected by the presence 
of impurity lattice sites.

\begin{figure}[!tbh]
\begin{tabular}{cc}
\centerline{
\resizebox{9.cm}{!}{%
\includegraphics[angle=-0]{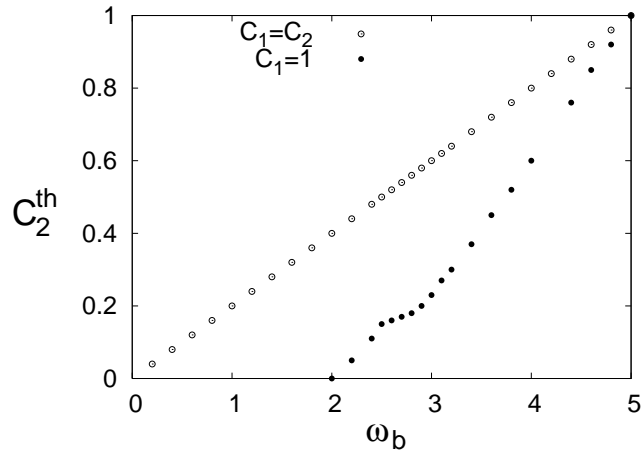}
} }
\end{tabular}
\caption{
Evolution of the threshold value of the coupling
parameter, $C_{2}^{th}$, as a function of the frequency, $\omega_{b}$,
for two different continuations starts. The values of $C_{2}^{th}$
limit the region where pinned discrete breathers are linearly
stable (unstable for $C_{2}>C_{2}^{th}$). The instability yields a
hyper-localized state (quasi-collapse). The continuation from the
fully uncoupled limit ($C_{1}=C_{2}=0$) (filled circles) is
performed using the path $C_{1}=C_{2}$. For the continuation (bold
circles) from the 1-dimensional limit ($C_{1}=1$, $C_{2}=0$) the
coupling in the new direction $C_{2}$ is progressively increased.
} 
\label{fig:Ctresholds}
\end{figure}

As expected, our results further corroborate the existence of
quasi-collapse instabilities in the anisotropic case: The phase
diagram in parameter space ($\omega_b$, $C_2$, $\nu$) consists of two
regions (stable and unstable) separated by the surface of
transition. As we perform the continuation of breather solutions
across  the parameter space we scan the Floquet stability of
the computed solution. In figure \ref{fig:Ctresholds} we present the two
stability transition curves in the plane ($\omega_{b}$, $C_{2}$,
$\nu=1$), {\em i.e.} the function $C_{2}^{th}(\omega_{b})$, 
corresponding to the two different continuation starts. The
continuation from the AC limit is made through the path $C_{1}=C_{2}$
and the one from the 1D limit is made at $C_{1}=1$. The convergence of
the two paths at $C_{2}=1$ is clearly seen. 

\begin{figure}[!tbh]
\begin{tabular}{cc}
\centerline{
\resizebox{14.cm}{!}{%
\includegraphics[angle=-0]{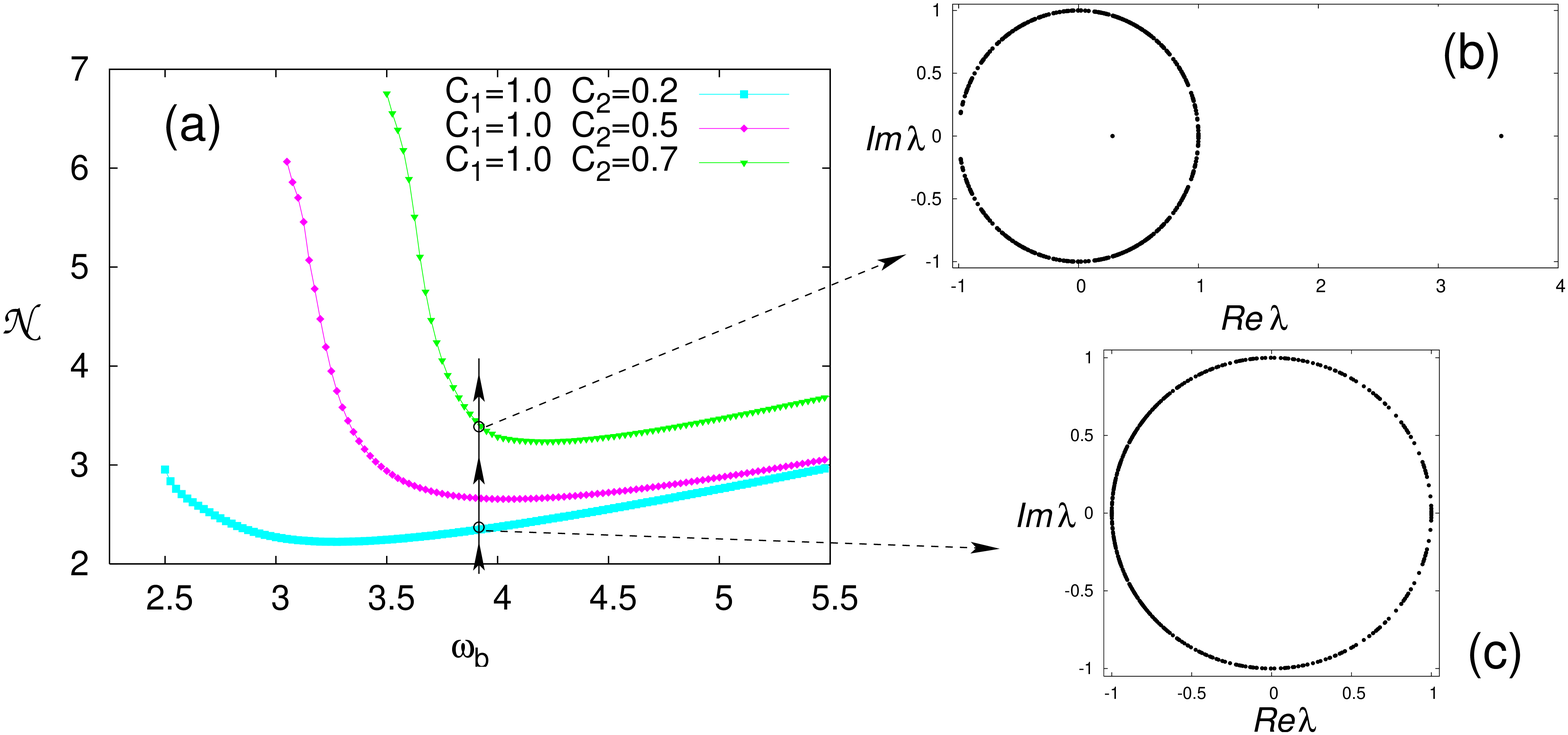}
} }
\end{tabular}
\caption{
{\bf (a)} Plot of the Norm, $\mathcal{N}$, of the computed
solutions as a function of their frequency $\omega_{b}$ for
different values of the coupling parameters. The continuations
have been made starting from the 1-dimensional limit ($C_{1}=1$).
For the regions where $\partial \mathcal{N}/\partial \omega_{b}$ is
positive (negative) the continued solutions are stable (unstable).
We can monitor the change of the linear stability of a solution of
a given frequency during its continuation in $C_{2}$ looking at
the Floquet spectra. Figures {\bf (b)} and {\bf (c)}  show the
Floquet spectra of a discrete breather of frequency
$\omega_{b}=3.93$  at $C_{1}=1$, $C_{2}=0.7$ (where $\partial
\mathcal{N}/\partial \omega_{b}<0$) and at $C_{1}=1$, $C_{2}=0.2$
(where $\partial \mathcal{N}/\partial \omega_{b}>0$), respectively.
} 
\label{fig:Norma-C2}
\end{figure}

\begin{figure}[!tbh]
\begin{tabular}{cc}
\centerline{
\resizebox{14.cm}{!}{%
\includegraphics[angle=-0]{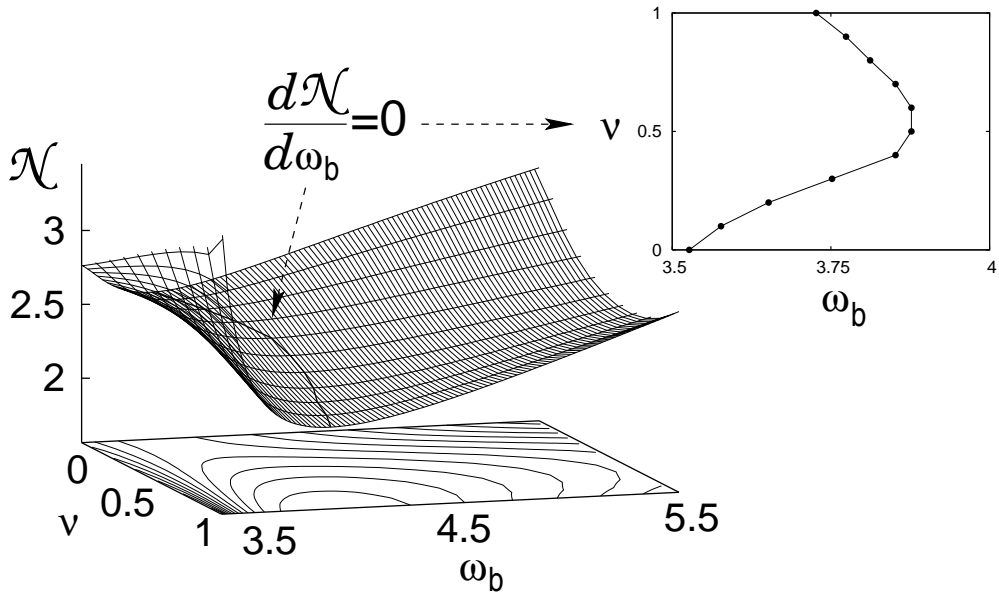}
} }
\end{tabular}
\caption{
Surface $\mathcal{N}(\omega_{b},\nu)$ for the case $C_{2}=0.5$. The inset 
shows the curve $\nu(\omega_{b})$ corresponding to $\partial \mathcal{N}
/ \partial \omega_{b}=0$. This curve gives the transition points where 
the discrete breather changes its stability character. 
} 
\label{fig:Norma-nu}
\end{figure}

The criterion for stability of the pinned discrete breather solution
derived in \cite{Laedke-PRL94,Laedke-JETP95},
\begin{equation}
\left(\frac{\partial \mathcal{N}}{\partial
\omega_b}\right)_{C_2,\nu} > 0\;\;,
\label{eq:stability}
\end{equation}
is of a very general character and our numerics illustrate it
clearly. On the other hand, the Floquet stability analysis detects 
the dimensionality (and a basis in tangent space) of the 
unstable linear manifold associated with the quasi-collapse instability 
that these exact discrete breathers experience for some parameter values.
We have computed numerically, for a fine grid of $\omega_b$ values 
and a coarser grid of $C_2$ and $\nu$, the function 
$\mathcal{N}(\omega_b, C_2, \nu)$, from which we show some sectors 
in figures \ref{fig:Norma-C2} and \ref{fig:Norma-nu}.

In figure \ref{fig:Norma-C2} we show the numerically computed 
norm (\ref{eq:2DNorm}) as a function of the breather frequency 
$\mathcal{N}(\omega_b)$, for three different values of the 
transversal coupling $C_2$, and a fixed value of $\nu = 1$ 
(anisotropic DNLS limit). We observe the existence 
of a minimum value, $\min \mathcal{N}(\omega_b)= 
\mathcal{N}^{th}\neq 0$, which is thus seen as an {\em excitation 
threshold} for the creation of these solutions. The position 
of the minimum $\omega^{th}_b(C_2)$, which {\em naturally} 
increases with $C_2$, separates the stable and unstable
branches of pinned breathers. Breathers corresponding to values of
$\omega_b$ where $\mathcal{N}(\omega_b)$ has a negative slope are
unstable: This is shown in the insets, where the Floquet spectra of
two representative examples of pinned discrete Schr\"odinger
breathers are plotted in the complex plane. Note that the high
accuracy of the numerical solution allows an unprecedented
detailed Floquet analysis of the instability, paving the way to
rigorous analytical characterizations of the quasi-collapse
unstable manifold. This is a one-dimensional manifold, as our
numerical results unambiguously confirm. Then, in the regime of
small time scales, the unstable manifold is fully characterized by
a single Floquet eigenvector.

Figure \ref{fig:Norma-nu} shows the (surface) function $\mathcal{N}(\omega_b, \nu)$
for the volume sector of constant $C_2$($= 0.5$). Most
noticeably, the critical (threshold) line of bifurcation points
($\frac{\partial \mathcal{N}}{\partial \omega_b} = 0$), as seen in
the inset, does not define a monotone function $\omega_b^{th}(\nu)$.
In fact, in the whole interval of $0 \leq \nu \leq 1$ values, the
range of values of $\omega_b^{th}$ is quite short,
indicating the insensitibity of the gross
features of the quasi-collapse transition to the value of $\nu$.
However, considering finer details, one sees that the threshold
curve $\omega_b^{th}(\nu)$ smoothly reaches its slightly larger
values around midway between the DNLS and the A-L
limits. In other words, intermediate values of the interpolating
(Salerno) parameter $\nu$ somewhat favour the enhancement of the
quasi-collapse unstable region. These conclusions are in contrast
with the stated conclusion (for isotropic lattices) in
\cite{Christ-Gaid-PRB96} that the Ablowitz-Ladik term increases
the stability regime.

When instability is allowed to develop beyond the fixed point
tangent space into the nonlinear realm of perturbations, the
trajectory obtained by direct integration of the equations of
motion invariably ends after a transient (of time scale given by
the real Floquet exponent larger than $1$) in a localized solution
with complex dynamics, the {\em pulson states}. These states
were characterized in \cite{Escorial} in the
following terms "... where the peak intensity $|\Phi_{m,n}|^{2}$
oscillates between the central site and its four nearest
neighbours... it is not known whether these pulson states
represent true quasiperiodic solutions to the DNLS equation". What
makes these trajectories on the unstable nonlinear quasi-collapse
manifold of much practical relevance and interest is their ubiquity:
They appear as persistent localized states in the hamiltonian
dynamical evolution from a wide variety of initial conditions. Their 
description requires at least two frequencies, namely the internal
(genuine breather-like frequency) and the frequency of the
oscillations of the {\em breather width} around a mean width
value, which turns out to be less than the width of the unstable
exact discrete breather. Second and outer shells
of neighbours (in both lattice axes) also participate in the
width oscillations.
\begin{figure}[!tbh]
\begin{tabular}{cc}
\centerline{
\resizebox{14.cm}{!}{%
\includegraphics[angle=-0]{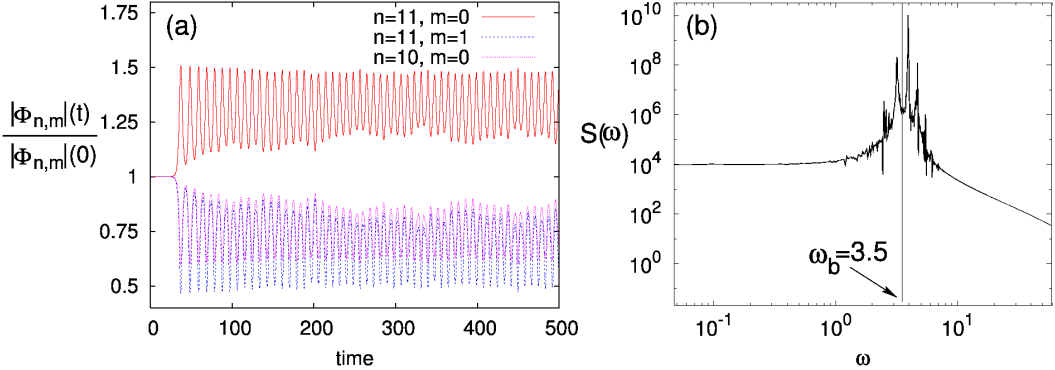}
} }
\end{tabular}
\caption{
{\bf (a)} Time evolution of the amplitude $|\Phi_{n,m}|$ for the 
localization center ($n=11$, $m=0$) and two adjacent sites ($n=10$, $m=0$)
and ($n=11$, $m=1$). Each amplitude is normalized to its initial
value, so that it can be seen how the quasi-collapse instability is
developed. The parameters of equation (\ref{eq:2D-DNLS}) are
$C_{1}=1$, $C_{2}=0.5$, $\nu=1.0$ ($\mu=0$) and the frequency of the
pinned breather is $\omega_{b}=3.50$. When the instability is fully
developed, we analyse the final state by means of the Power Spectrum 
$S(\omega)$ of the time evolution of
$\Re\left[{\Phi_{10,0}}(t)\right]$ (the real
part of localization center). 
As can be observed in {\bf (b)} the internal frequency of 
the breather (highest peak in the spectrum) shifts to a higher value 
($\omega^{*}_b=4.03$) and the other peaks are located at 
the frequencies of the harmonics resulting from the combination of the
internal frequency with the frequency ($\omega_{qc}=0.78$) asociated 
with the amplitude $|\Phi_{n,m}|$ oscillations shown in {\bf (a)}.     
} 
\label{fig:QC-FFT}
\end{figure}

Though a more detailed characterization of the pulson states will
be presented elsewhere, it is illustrative to consider 
(figure \ref{fig:QC-FFT}) the power spectrum of 
the field at the central site of a typical trajectory on 
the unstable nonlinear manifold of a quasi-collapsing 
pinned discrete breather. This shows peaks at the
combinations $\omega^{*}_b + j\omega_{qc}$ ($j=0, \pm 1, \pm
2...$), where $\omega_{qc}$ is the frequency of the width
oscillations characterizing the pulson state, while $\omega^{*}_b
> \omega_b$ is a frequency higher than the (initial condition)
fixed point frequency $\omega_b$. The new frequency $\omega^{*}_b$
turns out to be very close to the breather frequency of the same
(initial) norm on the stable branch. In other words, the
instability drives a shift of breathing frequency towards the
stable branch, while the excess energy is transferred to the
oscillatory motion of the observable width. This behavior 
seems to be the essence of the physical characterization 
of the nonlinear quasi-collapse manifold dynamics.

The numerical observation of a two-frequency power spectrum for a
typical pulson state points towards an eventual positive answer to
the question (on true quasiperiodicity) arised in \cite{Escorial},
though, for more details we have to refer to the PhD dissertation
\cite{JGGPhD}. This point serves to illustrate how the high
accuracy of the fixed point numerical solution provides detailed
clues on many still unsolved (from a mathematical and physical
point of view) questions on two-dimensional Schr\"odinger
localization, which are of prospective experimental
interest in nonlinear (photonic, Josephson, ...) physics
technologies.

\section{Breathers moving along the strong coupling direction.}
\label{secc:mobile}

Early and current attempts to explore straightaway discrete
breather mobility in isotropic 2D Schr\"odinger lattices seem to
agree \cite{Cuevas} that "kicking" procedures meet huge
difficulties in delivering good mobile solutions,
contrary to the numerical experiences in 1D lattices. We note 
here that the formal basis for those methods \cite{Chen} takes
advantage of the Floquet spectra analysis of exact pinned
breathers, where the so-called {\em depinning} (symmetry-breaking)
mode is identified. This allows, {\em provided Peierls-Nabarro barriers
are small enough}, to obtain nice numerical 1D mobile discrete
breathers, by computing trajectories from perturbations of the
exact pinned breather along the tangent space direction specified
by the depinning eigenvector. The presence of symmetry-breaking
instabilities leading to exchange of stability between one-site
and two-site centered pinned breathers
\cite{Tesis-Cretegny,dissipa1} and the associated lowering of the
Peierls-Nabarro barriers to breather displacements, hugely facilitating the
success of these procedures when applied to (both hamiltonian and
dissipative) one-dimensional lattices.
\cite{dissipative,Meister,Tesina-Oster}

In contrast, our "anisotropic lattice" continuation approach takes
advantage of the availability of exact 1D mobile solutions by
monitoring the parameter $C_2$ of transversal coupling, and then
does not rely on how easily one promotes clean mobility from
pinned localization. In this way we obtain accurate numerical
($p_x,p_y=0,q$) fixed points, that is Schr\"odinger discrete
breathers moving along the strong coupling direction. In a forthcoming
paper we will address the (much more difficult) question for
arbitrary direction of motion.

\subsection{Structure and stability of (1,0,1) fixed points.}
\label{subsecc:structure}

\begin{figure}[!tbh]
\begin{tabular}{cc}
\centerline{
\resizebox{14.cm}{!}{%
\includegraphics[angle=-0]{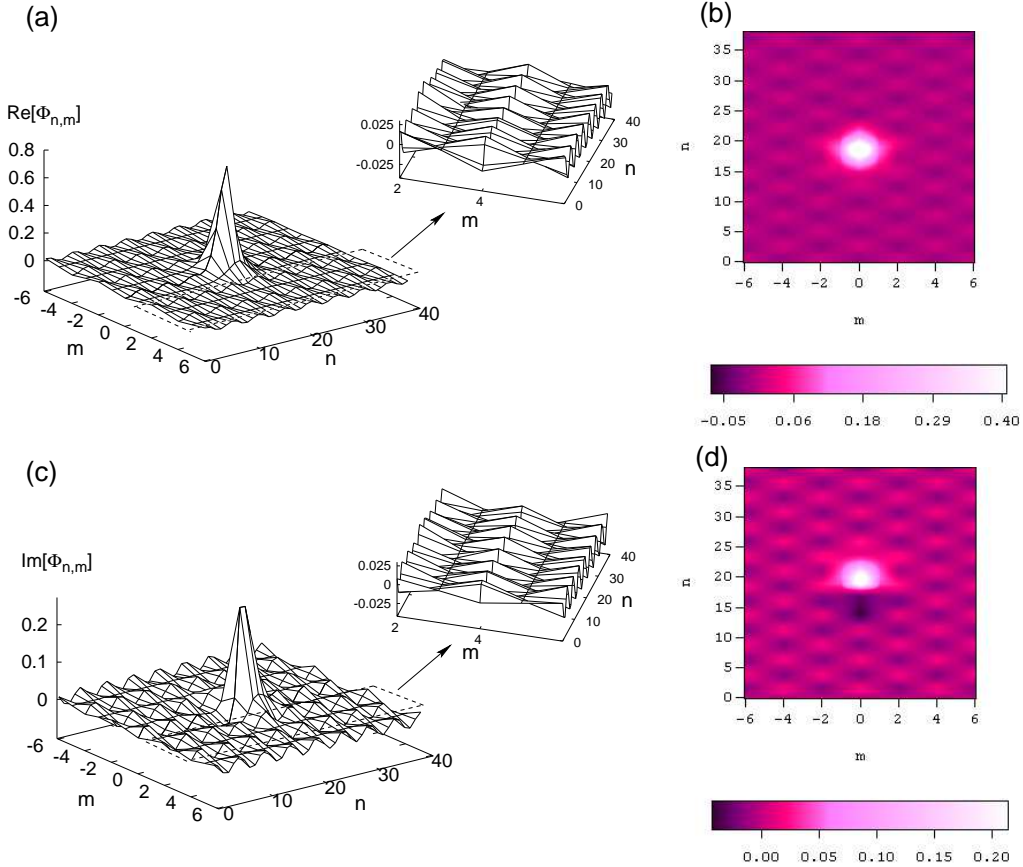}
} 
}
\end{tabular}
\caption{
Real, {\bf(a)} and {\bf (b)}, and imaginary, {\bf(c)} and {\bf (d)},
of a mobile ($1, 0, 1$)-discrete breather of frequency
$\omega_{b}=2.712$. The parameters of equation (\ref{eq:2D-DNLS}) are
$C_{1}=1$, $C_{2}=0.14$ and $\nu=0.95$ ($\mu=0.05$). The insets in 
{\bf (a)} and {\bf (c)} show the background far from the moving
core. It can be observed that the wavenumbers in the transversal
direction are $k_{y}=\pm\pi/2$. {\bf(b)} and {\bf (d)} show the
contour plot for both real and imaginary parts.  
} 
\label{fig:Mob-Structure}
\end{figure}

In figure \ref{fig:Mob-Structure} we visualize the 
instantaneous real and imaginary components of the 2D 
discrete field profile of a typical (1,0,1) 
Schr\"odinger breather. Its structure can be seen as the
natural extension to two-dimensional lattices of the structure of
mobile Schr\"odinger breathers analyzed in \cite{JGG1,JGG2}. The
numerical solution is spatially asymptotic to a {\em finely tuned}
small-amplitude extended (delocalized) radiation state
$\Phi_{m,n}^{{\mbox {bckg}}}(t)$ when $m, n \rightarrow \infty$.
The fixed point solution can be thus decomposed as

\begin{equation}
\Phi_{m,n}(t)=\Phi_{m,n}^{{\mbox {core}}}(t)+\Phi_{m,n}^{{\mbox
{bckg}}}(t)\;,
\end{equation}

which defines $\Phi_{m,n}^{{\mbox {core}}}(t)$, the spatially {\em
localized} component of the solution. It turns out that the
spatially {\em delocalized} component is a highly localized state
in the (continuum, in the thermodynamic limit) $k$-space of
wavevectors. More precisely, $\Phi_{m,n}^{{\mbox
{bckg}}}(t)$ is a finite linear combination of (1,0,1)
-resonant nonlinear ({\em i.e.} amplitude-dependent frequency
$\omega$) 2D planewaves. It can be said that, as might be expected,
1D Schr\"odinger breather mobility smoothly persists when
(strong $C_1$-coupling) 1D chains are coupled transversally.
Importantly, the numerical continuation for increasing values of
the transversal coupling $C_2$ proceeds far from the weak
coupling regime into where the genuine two-dimensional effects
start to be manifest, as we will see below.

Most noticeable, the SVD-regularized Newton procedure 
invariably selects the values $k_{y}=\pm\pi/2$ for all values of $C_2$ and 
$\nu$, and thus the values of $k_x$ for the 2D resonant planewave are 
independent of $C_2$ (so it remains equal to the $k$ values of the 
1D ($1,1$) fixed point for the uncoupled chain). The appearance of an
extended background modulation in the transversal direction of
$k_y=\pm\pi/2$ appears naturally as the {\em best} choice to
take advantage of approximately 1D breather propagation along strong
coupling direction, for it keeps the value of $k_x$ favoured by
the strong coupling $C_1$ value: Any other value of $k_y$ would
entail a different $k_x$ value. Note however that this provides only a
{\em plausibility argument} for the interpretation of the
numerical observation ($k_y=\pm\pi/2$). 

The high accuracy of the computed solutions allows a detailed
analysis of many issues concerning 2D Schr\"odinger breather exact
mobility along the strong coupling direction. We leave aside in
this paper many of them, and focus here on how the existence of
quasi-collapse instabilities of pinned Schr\"odinger breathers,
for increasing $C_2$-coupling values, influences the stability
properties of moving ($1,0,1$) breathers. In other words, 
we search here for genuine 2D effects on these "strong-coupling
-direction" (quasi-1D) moving breathers.

We have performed an exhaustive exploration of two sectors of the
parameter space ($C_2, \omega_b, \nu$), corresponding to the
breather frequency values $\omega_b = 2.5843$, and $\omega_b =
2.712$, by computing the continued ($1,0,1$) fixed point. 
These values of $\omega_b$ were chosen low enough to allow the 
analysis of pinned breather quasi-collapse effects on mobility, which 
occurs at relatively low values of $C_2^{th}(\nu)$ for these 
values of $\omega$ .

\begin{figure}[!tbh]
\begin{tabular}{cc}
\centerline{
\resizebox{14.cm}{!}{%
\includegraphics[angle=-0]{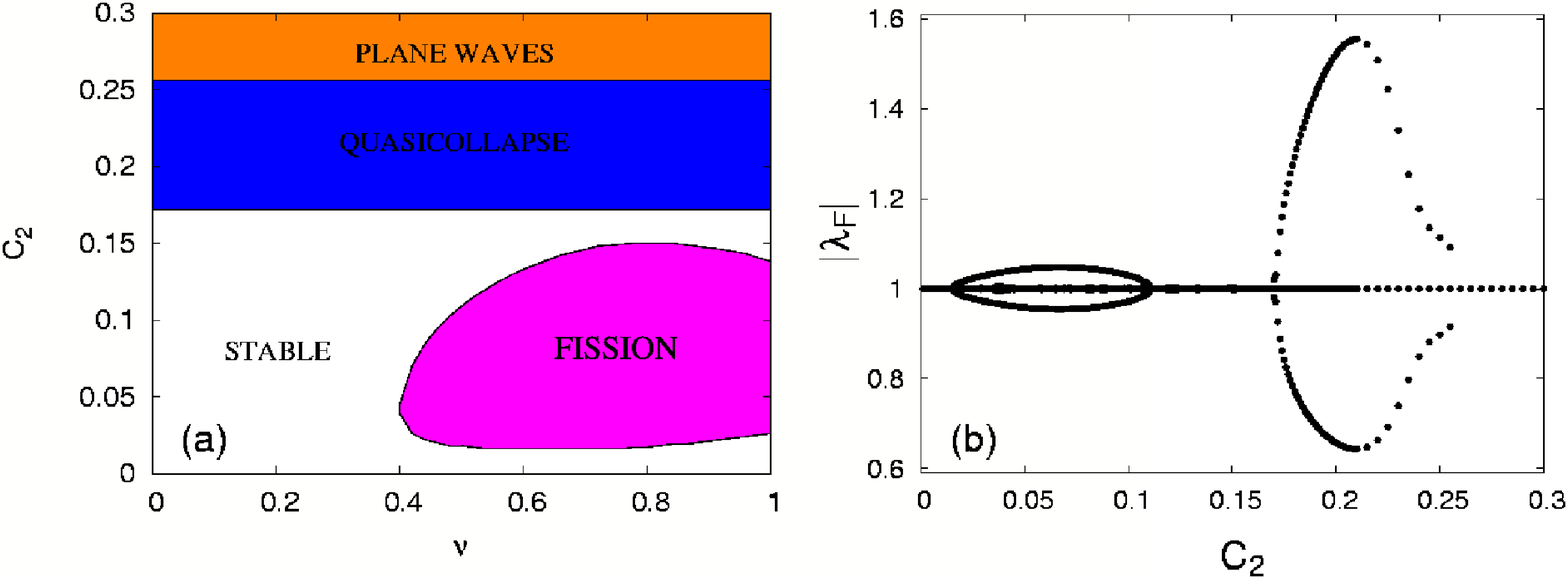}
} 
}
\\
\centerline{
\resizebox{14.cm}{!}{%
\includegraphics[angle=-0]{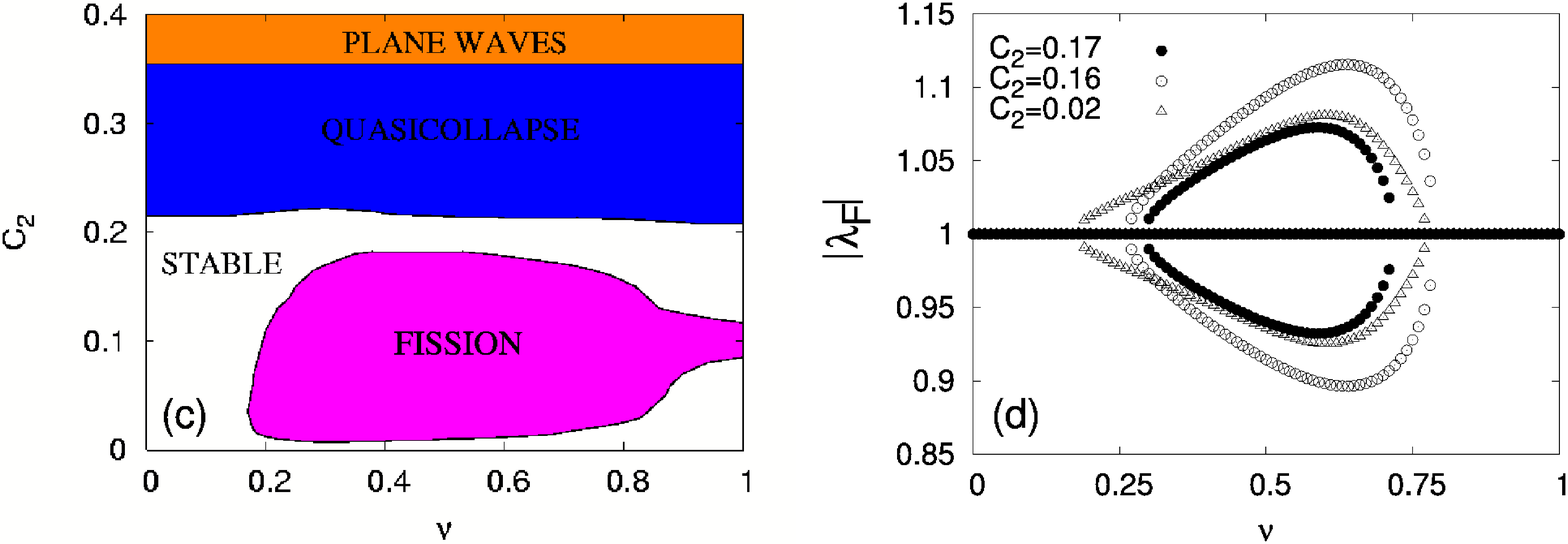}
} 
}
\end{tabular}
\caption{ 
Stability diagram and evolution of the modulus of the Floquet
eigenvalues for two ($1,0,1$)-discrete breather of 
frequencies $\omega_{b}=2.584$ {\bf (a)} and {\bf (b)}, 
and $\omega_{b}=2.712$ {\bf (c)} and {\bf (d)}. The
stability diagram {\bf (a)} and {\bf (c)} show two regions where the
mobile DB becomes unstable. For low values of the coupling $C_{2}$ there is a
subset of values of $\nu$ where the breather suffers from fission (see
text and figures (\ref{fig:UnstableDyn}.a) and
(\ref{fig:UnstableDyn}.b). On the other hand for higher values of
$C_{2}$ there is a second region (quasi independent of $\nu$) where
the unstable breather yiels a travelling quasi-collapsing state (see
text and figures (\ref{fig:UnstableDyn}.c) and
(\ref{fig:UnstableDyn}.d)). The evolution of the modulus of the Floquet
eigenvalues along different paths
$\nu=0.50$ {\bf (b)} and $C_{2}=0.17$, $0.16$, $0.02$ {\bf (c)} is
shown.   
} 
\label{fig:Mob-Stability}
\end{figure}

The Floquet analysis of the computed solutions provides the
stability diagrams represented in figures \ref{fig:Mob-Stability}. 
Both show no qualitative differences: There are two regions in the ($C_2, \nu$)
plane where the ($1,0,1$) mobile breather is linearly unstable.
The figures are not "schematic": Every point of the plane
in a fine grid of values of $C_2$ and $\nu$ has been analyzed,
{\em i.e.} the Floquet spectrum of the computed ($1,0,1$) fixed
point is scrutinized, as shown in figures \ref{fig:Mob-Stability}.b
and \ref{fig:Mob-Stability}.d, where the modulus of
the Floquet eigenvalues is shown as a function of either 
$\nu$ (figure \ref{fig:Mob-Stability}.b) or $C_2$ 
(figure \ref{fig:Mob-Stability}.d).

The first unstable region appears at low values of $C_2$ and
intermediate to high values of the Salerno parameter $\nu$, {\em
i.e.} it does not occur close to the A-L limit. This
unstable region is also bounded above in the direction of $C_2$:
The variation of the modulus of the unstable Floquet eigenvalue
versus the transversal coupling parameter $C_2$ shows that the
mobile breather becomes stable again at larger values of $C_2$,
before the second instability at even higher coupling takes
place. An important observation is that the pinned discrete
breather of the same frequency is {\em linearly stable} at the
points in this unstable region for (1,0,1) mobile breathers.
Thus this instability cannot be ascribed to pinned quasi-collapse
effects.

The second transition occurs for values of $C_2$ close to, but
slightly higher than, the values $C_2^{th}$ of the quasi-collapse
of the pinned breather of the same frequency. We had already seen
in the previous section that the quasi-collapse transition
$C_2^{th}(\nu)$ is only very weakly dependent on the value of
$\nu$, and note that the same is true for this mobile breather
bifurcation. These results suggest that this second transition is
related to quasi-collapsing phenomena. Significantly, the
stability of the ($1,0,1$) mobile breather persists for a small
interval of coupling values above the pinned breather
quasi-collapse. This should be regarded as natural, for the
mobile breather is a different solution. Note in figure 
\ref{fig:Mob-Stability}.b that the modulus of the unstable 
Floquet eigenvalue, in the interior of the unstable region, 
reaches much higher values than those typical for the first type 
of instability, and decreases for larger values of $C_2$, 
before the breather solution ceases to exist and only 
plane wave solutions are obtained by our numerical method. 
Note that this behaviour of the unstable Floquet eigenvalue also 
fits well to the main features of the pinned quasi-collapse 
instability strength, as described by the slope $\partial 
\mathcal{N} / \partial \omega_{b}$. From now on we will refer 
to this instability of mobile breathers as the quasi-collapse
instability.

In the next subsection we characterize both generic 
types of instability, by looking at the details of
the unstable manifold associated with each type. As we will see,
pulson states turn out to play a role in the description of
typical trajectories on the unstable nonlinear manifolds.

\subsection{Unstable manifold behaviour and ubiquity of pulson states.}
\label{subsecc:structure}

\begin{figure}[!tbh]1
\begin{tabular}{cc}
\centerline{
\resizebox{14.cm}{!}{%
\includegraphics[angle=-0]{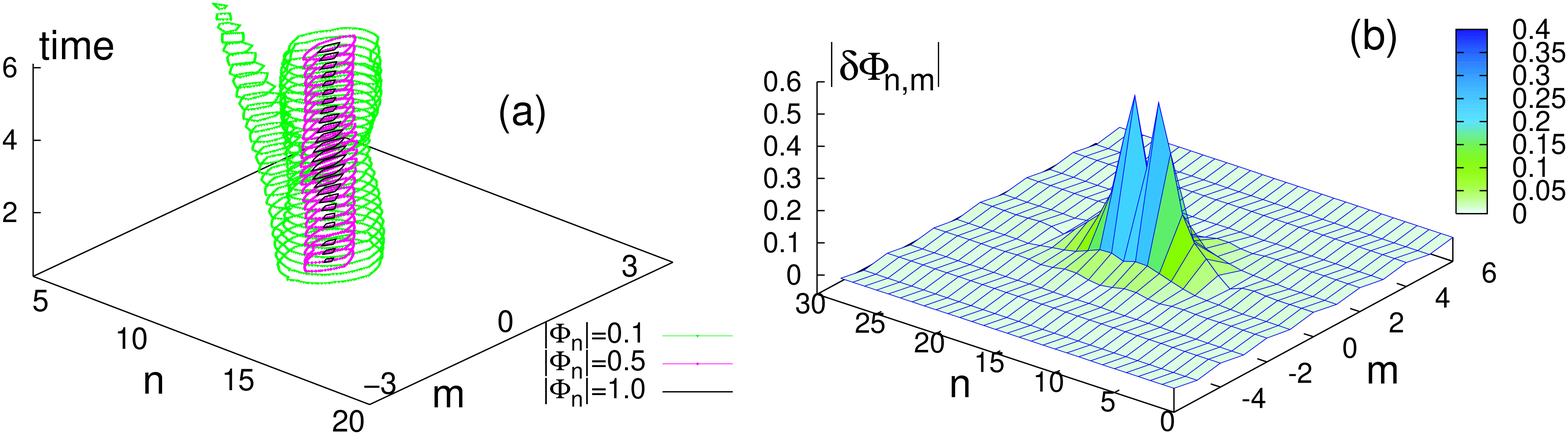}
} 
}
\\
\centerline{
\resizebox{14.cm}{!}{%
\includegraphics[angle=-0]{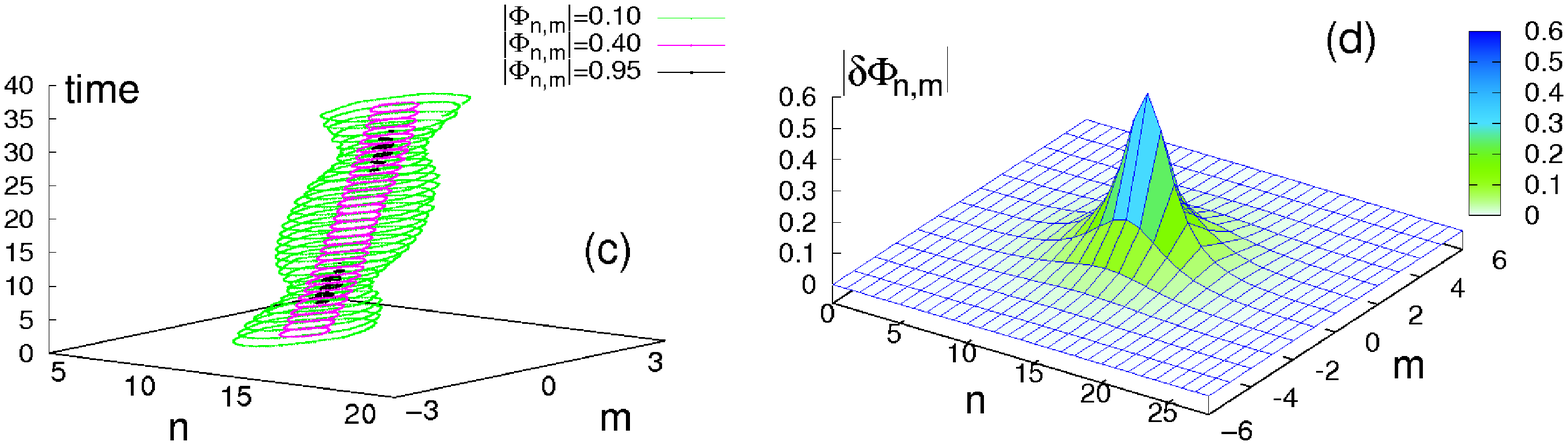}
}
}
\end{tabular}
\caption{
Time evolution of two unstable solutions, {\bf (a)} and {\bf
(c)}, of frequency $omega_{b}=2.584$ and the asociated unstable 
Floquet eigenvector, {\bf (b)} and {\bf (d)} respectively. 
Figures {\bf (a)} and {\bf (c)} show the time evolution of the 
contour lines correponding to three different values of
$|\Phi_{n,m}|$, in order to visualize the 4-dimensional
functions $|\Phi_{n,m}|(t)$. Figures {\bf (a)} and {\bf (b)} shows the
fission of the breather solution when perturbed along the
unstable ``M-shaped'' Floquet eigenvector plotted in {\bf (b)}. 
It can be seen how a low amplitude pulse emerges 
and the mobile breather becomes pinned. After this
transient this low amplitude  pulse decays into radiation. The parameter of
equation (\ref{eq:2D-DNLS}) are $C_{1}=1$, $C_{2}=0.08$ and 
$\nu=0.5$ ($\mu=0.5$). In the case of figures {\bf (c)} and {\bf (d)}
the parameters are the same except for $C_{2}=0.19$. In this case the
solution is in the ``quasi-collapse'' unstable region shown in figure
(\ref{fig:Mob-Stability}.a). The final state when perturbed along the
unstable eigenvector {\bf (d)} is a travelling breather whose
amplitude oscillates in the same fashion as that of the pinned
quasi-collapsing breathers, {\em i.e} the localization center
oscillates out of phase with repect to all the other sites on the lattice.     
} 
\label{fig:UnstableDyn}
\end{figure}

First, we analyze the quasi-collapse instability of ($1,0,1$)
mobile breathers. The unstable linear subspace in the tangent
space of the fixed point is one-dimensional. The typical
instantaneous profile of the (modulus) unstable Floquet
eigenvector driving the instability is shown in figure 
\ref{fig:UnstableDyn}.d. It is an exponentially localized 2D 
profile which decays asymptotically to zero as $m, n \rightarrow 
\infty$, {\em i.e.} it does not excite radiation. These 
characteristics are shared by the quasi-collapse unstable 
eigenvector of the pinned breathers, which further reinforce 
the previous considerations leading us to consider this instability 
as the mobile counterpart of the pinned quasi-collapse transition.

\begin{figure}[!tbh]
\begin{tabular}{cc}
\centerline{
\resizebox{14.cm}{!}{%
\includegraphics[angle=-0]{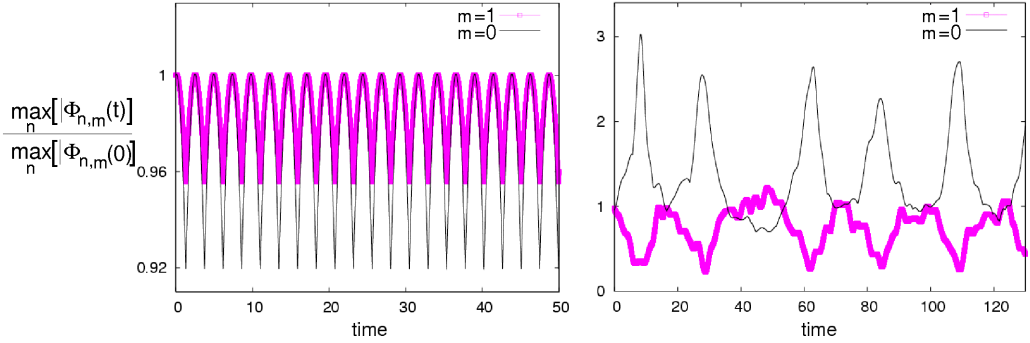}
} 
}
\end{tabular}
\caption{Time evolution of the maximum value of the modulus
$|\Phi_{n,m}|(t)$ along the central ($m=0$) chain and the adjacent ($m=1$)
one for a mobile ($1,0,1$) breather with frequency
$\omega_{b}=2.584$. This magnitude is normalized to the initial
value $|\Phi_{n,m}|(t_{0})$. Figure{\bf (a)} shows this evolution for an stable situation
($C_{2}=0.15$, $\nu=0.5$). It can be observed how the localization center
($m=0$) and is neighbour in the transversal direction ($m=1$) follows
two in-phase periodic trajectories in their modulus due to the
Peierls-Nabarro barrier surpassed during the motion. In contrast,
figure {\bf (b)}, shows the case when the breather is unstable
($C_{2}=0.19$, $\nu=0.5$). Here the quasi-collapse dynamics is manifested while
the localization center moves across the lattice. As can be observed,
the oscillations of the two amplitudes are out of phase and the
amplitudes of these oscillations are one order of magnitude higher 
than those of figure {\bf (a)}.  
 } 
\label{fig:QC-Evolution}
\end{figure}

In figure \ref{fig:UnstableDyn}.c we have visualized the time 
evolution of the field modulus contour plot for a typical trajectory on the unstable
manifold. This is obtained by direct numerical integration of the
equations of motion, from an initial condition in which a small
perturbation along the quasi-collapse eigenvector has been added
to the unstable fixed point solution. One sees that the breather
translational motion slows down, and the energy is transferred to
width oscillations. These oscillations turn out to be more
irregular, see figure \ref{fig:QC-Evolution}, 
than those observed in section \ref{sec:pinned} when 
we inspected typical trajectories on the unstable nonlinear 
manifold of pinned breathers.

The difference in the character of the width oscillations in both
(pinned and mobile) cases may be ascribed to the presence of an
extended background component in the mobile breather solution,
which naturally enters into the energy transfer taking place
during temporal evolution. The slowing down of the translational
motion continues and eventually the breather pins into a
convulsive pulson state surrounded by the remaining radiation.

Now we pay attention to the ``low $C_2$'' instability of ($1,0,1$)
mobile breathers. The modulus profile of the unstable Floquet
eigenvector that drives this instability is M-shaped (bimodal), as
shown in figure \ref{fig:UnstableDyn}.b, and is asymptotic to an extended
planewave-like profile as $m, n \rightarrow \infty$, {\em i.e.} it
is not a purely localized perturbation. It is indeed rather
different from the quasi-collapse unstable eigenvector analyzed
above, which is consistent with the fact that the pinned breather
of the same frequency is linearly stable in this region of
parameter space. As argued above, this instability is not
related to quasi-collapse phenomena, and it does not appear in the
region of small values of the Salerno parameter $\nu$, close to
the A-L limit.

A typical trajectory on the unstable manifold associated with this
instability is shown in figure \ref{fig:UnstableDyn}.a, where we have plotted the
time evolution of the field modulus contour plot. We can see there
that the mobile breather pins quickly while a small pulse moving
backwards is ejected, which spreads and finally mixes with the
remaining delocalized background. However some energy is
transferred to width oscillations of the pinned breather so that
also in this case we observe the formation of pulson states
surrounded by the remaining radiation. As the main difference of
this behaviour, with respect to the evolution observed on the
quasi-collapse unstable manifold, is the ejection of the small
moving pulse, we refer to this instability as {\em fission}.

By increasing the strength of the initial perturbation along the
direction of the unstable eigenvector, one observes that the size
of the ejected pulse increases. This observation is consistent
with the results reported in \cite{Christ-Gaid-PRB96}, where the
evolution of initial moving gaussian pulses in isotropic 2D
Schr\"odinger lattices was studied. These numerical
experiences lead the authors to conclude that "the
characteristic feature of the discrete quasi-collapse of a moving
pulse is the splitting of the initially moving broad pulse into a
track of the standing narrow structures ..." (sic). However, we see from
our study of the stability of exact moving discrete breathers that
the fission and the quasi-collapse instabilities have different
origins and they appear in different regions of parameter space.
On the other hand, the ubiquitous phenomenon of width oscillations
of pinned localized structures (pulson states) cannot be ascribed
to quasi-collapse. They also appear as the preferred way to
allocate excess of (localization) energy in regions of parameter
space far from the quasi-collapse unstable region.

\section{Conclusions and Prospective Remarks}
\label{secc:conclusions}

We have studied here the dynamics of exact numerical discrete
breathers, both pinned and mobile, in two-dimensional anisotropic
nonlinear Schr\"odinger lattices. These solutions are computed using a
SVD-regularized Newton method by continuation from a set of uncoupled
1D chains into increasing non-zero values of the coupling in the
transversal direction.

We have performed an extensive exploration in the parameter space
($\omega_{b}$, $C_{2}$, $\nu$) of breather frequency, 
transversal coupling and Salerno parameter, by computing 
the Floquet spectra of the numerical solutions. We have also
computed the breather norm ${\mathcal N}(\omega_{b},C_{2},\nu)$ 
and further corroborate the general validity of the criterion 
found in \cite{Laedke-JETP95}, namely that the 
partial derivative $\partial{\mathcal N}/\partial \omega_{b}$ 
is positive for stable pinned breathers. Furthermore, we
have analyzed the dynamics on the quasi-collapse unstable manifold, where the
unstable breather experiences a shift in frequency towards the
(higher) value of the stable breather with the same norm. The excess
of energy is coherently transferred to oscillations of the breather width,
so that the resulting pulson state is characterized by two frequencies. 

We have studied discrete breathers moving along the strong coupling 
direction. These solutions are composed of an exponentially localized core 
on top of an extended background which is itself the finite sum of a finite set of 
nonlinear 2D plane waves. The time scales asociated with these plane 
waves are resonant with the core internal frequency as happens in the 
1D case. In particular, the background chooses a finite set of plane 
waves from a continuous family of resonant solutions. The Floquet
analysis of these mobile discrete breathers reveals the existence of
two distinct types of instability. One is the counterpart, for mobile
breathers, of the quasi-collapse experienced by pinned breathers. The
other instability occurs in a region of parameter space where pinned
breathers are linearly stable. The analysis of the dynamics on the
unstable manifold show that the excess of energy is partly transferred
to a small moving pulse, ejected from the center of localization,
which justifies the designation of a fission instability. However, part of the
energy excess is also transferred to width oscillations. The
appearance of pulson states far from the quasi-collapse regime
indicates that the tendency to allocate energy in the form of width
oscillations is a general 2D feature, not exclusively associated to
quasi-collapse instabilities.

In a future work we will focus on mobility of 2D discrete breathers in 
an arbitrary lattice direction. The results obtained here shed light 
about how this mobility can be obtained. In fact, our experiences 
show that mobility of pinned breathers can be induced based 
on the existence of the extended background in the numerically 
exact mobile solution. On the other hand, the results obtained 
here and the aforementioned future work may help to design and 
better understand recent numerical experiments reported in
\cite{Eugenieva}, concerning the interaction between high amplitude 
pinned breathers and mobile ones. These experiments provides a
possible way for routing and blocking mobile discrete breathers 
via the interaction with the high amplitude pinned ones, resulting 
in a plausible implementation of logical functions. 
         
\section{Acknowledgments} 

The authors acknowledge F. Falo, Yu. Kivshar, R.S. Mackay and 
M. Peyrard for sharing thoughts, and pointing out some important 
references to us. JG-G and LMF are grateful to M. Johansson and 
B. Malomed for discussions on some issues regarding the 
"travelling wave" (orthodox) perspective on discrete breathers. 
Financial support came from MCyT (Project No. BFM2002 00113), DGA 
and BIFI. JG-G akcnowledges financial support from the MECyD through
a FPU grant. Work at Los Alamos performed under the auspices of the 
US DoE.

\end{document}